\documentclass[prd,twocolumn,nofootinbib]{revtex4}
\usepackage{bm}
\usepackage{amssymb}
\usepackage{graphicx}
\usepackage{epstopdf}
\usepackage{mathrsfs}
\usepackage{amsmath}

\usepackage[utf8]{inputenc}
\usepackage[english]{babel}
\usepackage{xcolor}

\begin{document}
\title{Climbing up the memory staircase: Equatorial zoom-whirl orbits} 
\author{Lior M.~Burko$^1$ and Gaurav Khanna$^2$
}
\affiliation{
$^1$ Theiss Research, La Jolla, California 92037, USA \\ 
$^2$ Department of Physics, University of Massachusetts, Dartmouth, Massachusetts  02747, USA
}
\date{July 24, 2020}
\begin{abstract} 

The nonlinear, or null gravitational memory effect for zoom-whirl orbits around fast spinning Kerr black holes for extreme mass ratio systems has a staircase structure. The latter is characterized by a periodic fast increase in the memory during the whirl phases of the orbit, and near  constant levels when the orbit is near apoapsis. We study the contribution of different spherical harmonic modes to this effect, and discuss the relative importance thereof. Because the hereditary memory effect increases in magnitude linearly in the number of periapsis passages on a dynamical time scale, while the oscillatory parent gravitational waves are sourced by an orbit that evolves on a much longer radiation reaction time scale, the magnitude of the former relative to the latter increases with time. We also consider orphan memory and  the memory effect for extreme mass ratio inspirals. We then discuss the prospect of detection with the Laser Interferometer Space Antenna (LISA). 

\end{abstract}
\maketitle


\section {Introduction}

After a gravitational wave passes through a freely falling detector, the latter does not return to its former configuration, but instead maintains a persistent memory of the gravitational wave.  There are two kinds of gravitational memory effects, arising from (i) waves that do not reach future null infinity ${\mathscr{I}^+}$ (regular memory, linear memory), and which arise from a change in the second time derivative of the quadrupole moment of the source, e.g., in the gravitational scattering of compact objects \cite{newman,zeldovich, braginsky86,braginsky87}, and (ii) waves that reach future null infinity ${\mathscr{I}^+}$ (null memory, nonlinear memory, Christodoulou memory), which are caused by the effective stress-energy of gravitational waves \cite{christodoulou,blanchet} (or that of massless particles such as neutrinos \cite{epstein78,turner78}) acting as a source for additional gravitational waves. 

The null memory effect is interesting from several points of view, specifically the nonlinear property of being gravitational waves sourced by gravitational waves, being a non-oscillatory effect, and being a hereditary effect that depends on the entire past history of the source. The memory effect can also provide information on the source that adds to the information obtained from the oscillatory parent gravitational waves, and can also add to the understanding of symmetries in General Relativity \cite{strominger}. Nevertheless, the gravitational memory effect has not been observed yet in gravitational wave experiments. 

A number of factors limit the prospects for the detection of the memory effect with terrestrial detectors, specifically the low frequencies associated with the characteristic rise time of the memory,  and the forces that push the proof masses of terrestrial interferometric detectors back to their equilibrium position after the wave has passed \cite{favata2010} (see, however, \cite{lasky}.)  Space-borne freely falling detectors such as the Laser Interferometer Space Antenna (LISA) do not share these limitations and therefore provide an interesting laboratory for the possible detection of gravitational wave memory. The different types of memory effect observables are listed in \cite{flanagan}. 

Because the magnitude of the memory effect for a single binary coalescence (such as GW150914) is rather small, and much smaller than the oscillatory parent gravitational wave, it is unlikely to be found in an experiment  \cite{lasky,johnson19}. A single merger can potentially be detected with LISA for coalescing two supermassive black holes \cite{islo19}. However, the prospects of detecting the memory effect become much better when considering a population of mergers, because the signal-to-noise ratio increases with the square root of the population size \cite{lasky}. 

A natural laboratory for measuring the regular memory effect for a different LISA source is the zoom-whirl orbits of extreme mass ratio systems \cite{glampedakis2002}. Zoom-whirl orbits are high eccentricity orbits around a rapidly rotating Kerr black hole, such that the semilatus rectum $p$ is close in value  to that of  the separatrix of stable bound orbits $p_{\rm s}$, i.e., $p\approx p_{\rm s}$. In zoom-whirl orbits the smaller compact object in an extreme mass ratio binary system ``zooms in'' from apoapsis, and near periapsis it completes a number of circular orbits, also known as whirls,  before ``zooming out" again towards apoapsis. 

When the compact object is in the whirl phase of its motion it emits strong gravitational waves, and specifically strong memory signals. As it moves in the zoom phases, and specifically close to apoapsis, the emission of either is much weaker because of the high eccentricity of the orbit. The hereditary memory effect therefore, when plotted as a function of time at a distant detector, has an appearance reminiscent of that of a staircase:  When the smaller member of the binary is far from periapsis there is little contribution to the memory effect, and the memory retains a roughly constant, or flat value.  When it is in the whirl phase of the orbit there is a rapid increase in the memory, analogous to the vertical rise part of a staircase. As the system continues to move more and more steps of the staircase are added to the signal.

The memory effect of zoom-whirl orbits continues to increase in amplitude with each periapsis passage. On the other hand, the oscillatory parent  gravitational waves do not (on a dynamical time scale, which is much shorter than the radiation reaction time scale). 
The memory effect becomes gradually more and more important, as its ratio to the oscillatory gravitational waves increases. This effect makes the memory of zoom-whirl orbits of interest in the search for the observation of the memory. 

A memory effect with a shape of a staircase resembles a series of step functions when the rise time is short. For a single short burst of gravitational waves the memory effect can be approximated by a Heaviside step function. 
In that limit the amplitude spectral density is inversely proportional to the frequency $f$ \cite{islo19,divarkala-2020}, and the power spectral density of the memory signal is inversely proportional to $f^2$. The memory effect is a low frequency effect, and therefore its relative contribution is greater at lower frequencies than at higher frequencies. This property may play a role in its identification and measurement.  
The characteristic strain is consequently independent of the frequency. With finite rise times the memory characteristic strain depends only weakly on the frequency. 

For zoom-whirl orbits the staircase-like signal, being the sum of approximate step functions, also has amplitude spectral density  proportional to $f^{-1}$, the power spectral density is proportional to $N^2/f^2$, where $N$ is the number of periapsis passages, and the characteristic strain is again independent of the frequency and is proportional to $N$, in  the low frequency approximation. When the characteristic strain of the oscillatory parent wave has higher  frequencies (say, because of lower masses) and lies outside the good sensitivity band of the detector, the memory remains mostly unchanged, and could become an orphan memory: a detectable memory signal without the associated parent wave \cite{mcneill_2017}.

The organization of this Paper is as follows. In Section \ref{sec:model} we describe the model we use to approximate the computation of the null  memory effect, and  in Section \ref{sec:num} we describe our numerical approach. We discuss our results for zoom-whirl orbits in Section \ref{sec:zw}, where we first discuss the quadrupole modes, then azimuthal modes. We next compare the magnitude of different mode, and finally orphan memory. We discuss inspiral orbits in Section \ref{sec:inspiral}, and make concluding remarks in Section \ref{sec:conc}.

\section{Modeling the Memory Effect}\label{sec:model} 
A number of approximations have been offered to model the memory effect. In most cases there are small differences in the amplitude of the memory effect, depending on the choice of the model.  Common models that appear in the literature are (i) the quadrupole approximation, which only considers the $\ell,m=2,2$ spherical harmonic mode \cite{favata2009}; (ii) the minimal waveform model, which is an analytical approximation based on the post-Newtonian expansion and the effective-one-body model \cite{favata2009}; (iii) the higher mode model, in which spherical harmonic modes up to $\ell=4$ are included \cite{talbot2018}; and (iv) the quadrupole kludge model, which uses detector-adapted coordinates to simplify angular integrals, and makes an assumption about the Isaacson stress-energy \cite{johnson19}. It was shown in \cite{boersma20} that the kludge approach leads in the quadrupole approximation (the $\ell,m=2,2$ mode) to a memory amplitude which is roughly half of that which is obtained in the other models, and contained a small oscillatory term. 

In view of the simplicity of the integration to find the memory effect with the kludge model, and in order to show the potential importance of the memory effect for zoom-whirl orbits, we model the memory effect using the kludge model. For more precise predictions one could use, say, the model presented in \cite{talbot2018}. Specifically, following \cite{johnson19} we use
\begin{eqnarray}\label{model}
h^{\rm{(mem)}}_{\ell,m}(u)&:=&h^{\rm{(mem)}}_{+\;\ell,m}(u)+h^{\rm{(mem)}}_{\times\;\ell,m}(u)\nonumber\\
&=&\frac{r}{68\pi}\Phi(\iota)\int_{-\infty}^{u}\,du' \langle \dot{h}_{+\;\ell,m}^2+\dot{h}_{\times\;\ell,m}^2\rangle\, ,
\end{eqnarray}
where $\Phi(\iota):=\,\sin^2\iota\,(17+\,\cos^2\iota)$, $u$ is retarded time, and an overdot denotes differentiation with respect to $u'$. The angle $\iota$ is the angle between the angular momentum vector of the binary and the line of sight between the binary and the observer. 
To maximize the effect we use $\iota=\pi/2$. The averaging in the integrand is to be performed over wavelengths. Following \cite{johnson19} we do not average twice, and ignore the averaging over the frequencies. It was shown in \cite{favata2011} that the memory effect depends only weakly on the frequency averaging, and when the frequency averaging is not done small amplitude oscillations at the orbital period are superimposed on the memory. We ignore such oscillations here given their smallness. We note that in our kludge model we do not make the assumption of the quadrupole approximation as is done in \cite{johnson19}, and compute the various $\ell,m$ modes. We only consider here positive azimuthal modes ($m\ge 0$) because the negative counterparts are calculable from the positive ones.


\section{Numerical approach}\label{sec:num}

We consider a fast spinning Kerr black hole of mass $M$ and spin angular momentum per unit mass $a$, and a compact object of mass $\mathfrak{m}$. In the extreme mass ratio case, $\mu:=\mathfrak{m}/M\ll 1$, the system's dynamics can be analyzed using the point-particle black hole perturbation theory. That is, the smaller compact object $\mathfrak{m}$ is modeled as a point particle with no internal structure, moving in the spacetime of the larger black hole $M$. Gravitational radiation is computed by evolving the perturbations of the Teukolsky function $\Psi$ which are generated by the moving particle, by solving the Teukolsky master equation with particle-source, which is written in Boyer-Lindquist coordinates as~\cite{emri_code}  
\begin{eqnarray}
\label{teuk0}
&&
-\left[\frac{(r^2 + a^2)^2 }{\Delta}-a^2\sin^2\theta\right]
        \partial_{tt}\Psi
-\frac{4 M a r}{\Delta}
        \partial_{t\phi}\Psi \nonumber \\
&&- 2s\left[r-\frac{M(r^2-a^2)}{\Delta}+ia\cos\theta\right]
        \partial_t\Psi\nonumber\\  
&&
+\,\Delta^{-s}\partial_r\left(\Delta^{s+1}\partial_r\Psi\right)
+\frac{1}{\sin\theta}\partial_\theta
\left(\sin\theta\partial_\theta\Psi\right)+\nonumber\\
&& \left[\frac{1}{\sin^2\theta}-\frac{a^2}{\Delta}\right] 
\partial_{\phi\phi}\Psi +\, 2s \left[\frac{a (r-M)}{\Delta} 
+ \frac{i \cos\theta}{\sin^2\theta}\right] \partial_\phi\Psi  \nonumber\\
&&- \left(s^2 \cot^2\theta - s \right) \Psi = -4\pi\left(r^2+a^2\cos^2\theta\right)T  \,  ,
\end{eqnarray}
where $\Delta = r^2 - 2 M r + a^2$ and $s$ is the ``spin weight'' 
of the field. The $s=-2$ case for $\Psi$ describes the radiative degrees of freedom of the gravitational field, the Weyl scalar $\psi_4$, in the radiation 
zone, and is directly related to the Weyl curvature scalar as $\Psi = (r - ia\cos\theta)^4\psi_4$. The Weyl scalar $\psi_4$ can then be integrated twice at future null infinity ${\mathscr{I}^+}$ to find the two polarization states $h_{+}$ and $h_{\times}$ of the transverse-traceless metric perturbations, 
\begin{equation}
\psi_4\approx\frac{1}{2}\left(\frac{\,\partial^2h_{+}}{\,\partial t^2}-i\frac{\,\partial^2h_{\times}}{\,\partial t^2}\right)\, .
\end{equation}

The source term $T$ in Eq.~(\ref{teuk0}) for the  smaller compact object $\mathfrak{m}$ is related to the energy-momentum tensor $T_{\alpha\beta}$ of a point particle, 
$$T_{\alpha\beta}=\mathfrak{m}\,\frac{u_{\alpha}u_{\beta}}{\Sigma\, \dot{t}\,\sin\theta}\delta[r-r(t)]\,\delta[\theta-\theta(t)]\,\delta[\phi-\phi(t)]\, ,$$
where $\Sigma=r^2+a^2\,\cos^2\theta$. Here, $\dot{t}:=\,dt/\,d\tau$, where $\tau$ is proper time, and $u^{\alpha}$ is the compact object's four-velocity. 
The source term $T$ is constructed by projecting $T_{\alpha\beta}$ onto the Kinnersley tetrad and operating on it with a certain second-order differential operator. For details see \cite{emri_code}.

Finding the  Weyl curvature scalar $\psi_4$ from the motion of $\mathfrak{m}$ around a Kerr 
black hole involves a two-step process. First, we compute the trajectory taken by the point-particle, 
and then we use that trajectory to compute the gravitational wave emission. For the first step, the 
particle's motion can simply be chosen to lie on a  geodesic for zoom-whirl orbits (see below in Section \ref{sec:zw}). The geodesic orbit is justified by the very different time scales of the compact object's motion, specifically the dynamical time scale being much shorter than the radiation reaction time scale, and the integration time being only a few dynamical time scales. In particular, we use 
the parameters $(p/M, e) = (2.250\,054\,2, 0.96)$ unless stated otherwise, with various values for $a/M$, where the orbital parameters are $p$, the semilatus rectum and  $e$, the eccentricity. These are the same orbital parameters used in \cite{wiggles}. 
For inspiral orbits, Section \ref{sec:inspiral}, we consider quasicircular orbits, i.e., orbits with vanishing eccentricity $e$ that would be circular in the absence of radiation reaction, and use balance arguments for otherwise conserved quantities to continuously adjust the orbit. For the second step, we solve the inhomogeneous Teukolsky equation in the time-domain while feeding the trajectory information from the first step into the particle source-term of the equation. 

We solve the Teukolsky equation (\ref{teuk0}) numerically in the time domain in a similar approach to that presented in \cite{emri_code}. In particular, (i) we first rewrite the Teukolsky 
equation using compactified hyperboloidal coordinates that allow us to extract the evolving fields 
directly at ${\mathscr{I}^+}$ while also solving the problem of artificial reflections from the outer boundary; 
(ii) we take advantage of axisymmetry of the background Kerr space-time, and separate the dependence 
on azimuthal coordinate, thus obtaining a set of (2+1) dimensional partial differential equations; 
(iii) we then rewrite these equations into a first-order, hyperbolic system; and in the last step (iv) 
we implement a two-step, second-order Lax-Wendroff, time-explicit, finite-difference numerical evolution 
scheme. The particle-source term on the right-hand-side of the Teukolsky equation requires some 
specialized techniques for such a finite-difference numerical implementation. Additional details can be 
found in \cite{emri_code} and the references included therein. The entire second step 
of the computation mentioned above is implemented using OpenCL/CUDA-based GPGPU-computing 
which allows for the possibility of performing very long duration and high-accuracy computations within 
a reasonable time frame. Numerical errors in these computations are typically on the scale of a small 
fraction of a percent~\cite{xsede_paper}. This numerical approach was used in \cite{wiggles}.

We present results for the waveforms and for the memory effect extracted at future null infinity ${\mathscr{I}^+}$ and along ${\mathscr{I}^+}$ as functions of retarded time, which we denote by $v$. In order to show the connection of the waves at ${\mathscr{I}^+}$  with the various parts of the orbit of the compact object, we parametrize the latter with retarded time $v$ using outgoing geometrical null rays. Specifically, we present below figures with the radial and angular Boyer-Lindquest coordinates $r,\varphi$ respectively, as functions of $v$ along the compact object's orbit, instead of parametrizing the parts of the orbit with the compact object's coordinate time $t$ or proper time $\tau$ to avoid retardation effects. 

\section{Zoom-Whirl orbits}\label{sec:zw}

\subsection{Quadrupole mode}

\subsubsection{Relative importance of the two polarization states}

We take the parameters of the zoom-whirl equatorial orbit to be $(p/M,e) = (2.250\,054\,2, 0.96)$ throughout, unless stated otherwise, where $M$ is the central black hole's mass, $p$ is the semilatus rectum and $e$ the eccentricity of the orbit. Here, we take $a/M=0.999\,9$. 
We first consider the quadrupole, $\ell,m=2,2$ mode.

\begin{figure}[t]
\includegraphics[width=8.5cm]{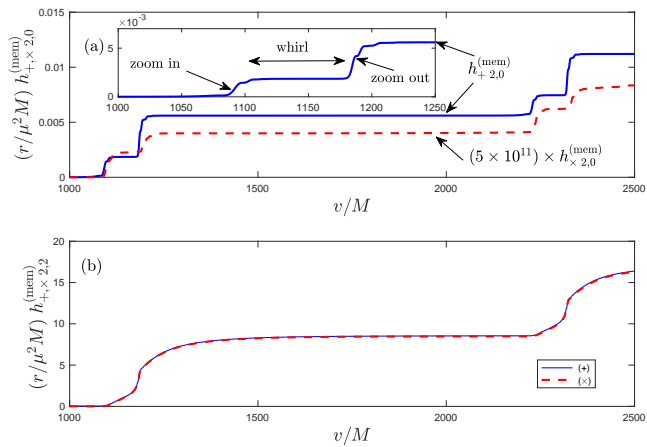}
\caption{The memory effect for both the $+$ and the $\times$ polarization states for zoom-whirl orbit for the $\ell,m=2,0$ and for the quadrupole mode ($\ell,m=2,2$). Upper panel (a): The $\ell,m=2,0$ mode. Notice that the magnitude of the $\times$ polarization is multiplied by $5\times 10^{11}$ to show detail on the same scale. The inset zooms into the transitional part of the curve for the $+$ polarization. Lower panel (b): The quadrupole mode ($\ell,m=2,2$). Notice that on the scale of this figure the $+$ and $\times$ polarizations are indistinguishable.}
\label{pluscross}
\end{figure}

We consider the question of the relative importance of the two polarization states in Fig.~\ref{pluscross}. The lower panel (b) shows the memory effect $h^{\rm{(mem)}}_{+,\times\;2,2}$ as a function of retarded time at the detector. 
The two polarization states of the memory effect are indistinguishable on the scale of this figure. The $+$ and $\times$ polarization states contribute comparably to the total effect. We note that it is possible to make a gauge choice when only the quadrupole mode is considered such that the memory from the $\times$ polarization state vanishes \cite{talbot2018}. We do not make this gauge choice here in order to compare the relative contributions of different modes in our approach. 

\subsubsection{Structure of the memory signal}

\begin{figure}[t]
\includegraphics[width=8.5cm]{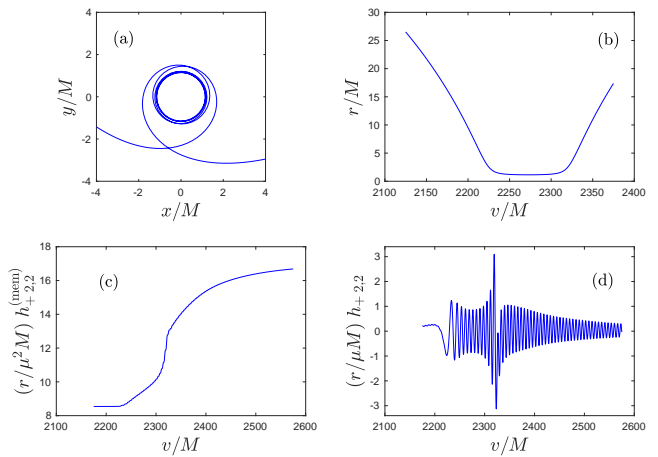}
\caption{Zoom-whirl orbit for the quadrupole mode ($\ell,m=2,2$). Panel (a): The inner part of the orbit in the $x-y$ plane. Panel (b): The coordinate $r$ as a function of retarded time $v$. Panel (c): The memory effect for the $+$ polarization state. Panel (d): The waveform. Panels (c) and (d) show the corresponding quantities for the same range of retarded time $v$. Panels (a) and (b) show a smaller range in $v$ to focus on the transition to and from the whirls.}
\label{22detail}
\end{figure}

We consider the same orbit in greater detail in Fig.~\ref{22detail}, which shows the shape of the orbit (Fig.~\ref{22detail}a), and the radial coordinate $r$   (Fig.~\ref{22detail}b), the memory effect (Fig.~\ref{22detail}c), and the oscillatory parent waveform (Fig.~\ref{22detail}d), as  functions of retarded time $v$. 

The rapid increase in $h^{\rm{(mem)}}_{+\;2,2}$ occurs during the whirl phase of the orbit. The exit from the whirl phase is accompanied with a greater increase in the memory effect than the entry into the whirl phase, because there are greater excitations of the black hole when the compact object is in the strong field. These excitations are seen in the waveform (Fig.~\ref{22detail}d). This asymmetry of the in and out parts of the zoom motion is another aspect of the asymmetry between them, that was noted in \cite{burko-khanna-2007} (see also \cite{wiggles}). 

\subsubsection{Effect of the central black hole's spin}

\begin{figure}[t]
\includegraphics[width=9.5cm]{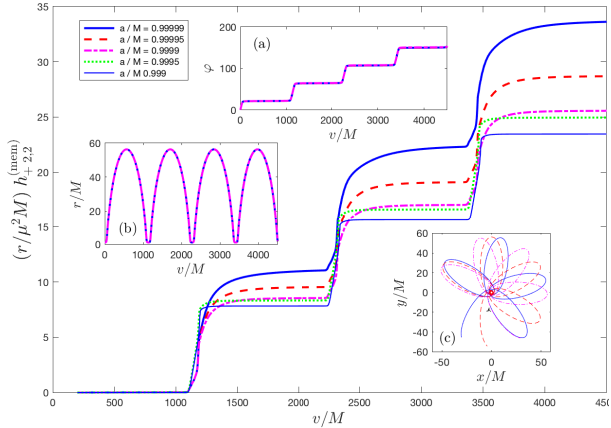}
\caption{The  memory effect $h^{\rm{(mem)}}_{+\;2,2}$ as a function of retarded time $v$ for a set of $a/M$ values. Bold solid curve: $a/M=0.999\,99$, dashed curve: $a/M=0.999\,95$, dash-dotted curve: $a/M=0.999\,9$, dotted curve: $a/M=0.999\,5$, and thin solid curve: $a/M=0.999$. The insets show (a) the angular coordinate $\varphi$ and (b) the radial coordinate $r$ as functions of $v$, and (c) the orbital shape in the $x-y$ plane,  for the cases $a/M=0.999\,9$, $0.999\,95$, and $0.999\,99$. }
\label{spin}
\end{figure}

The effect of the spin of the central black hole is shown in Fig.~\ref{spin}, which shows the memory effect $h^{\rm{(mem)}}_{+\;2,2}$ as a function of retarded time $v$ for a number of orbits that differ only in the value of $a/M$, but have the same orbital parameters $p/M,e$. The memory effect increases with the value of $a/M$. Specifically, we show the cases of $a/M=0.999, 0.999\,5, 0.999\,9, 0.999\,95$ and $0.999\,99$. 
The difference between the orbits is shown in the insets, which display  the angular coordinate $\varphi$ and the radial coordinate $r$ as functions of retarded time $v$ in panels Fig.~\ref{spin}a and Fig.~\ref{spin}b, respectively. The small difference in $\varphi$ which cannot be seen on the scale of Fig.~\ref{spin}b is manifested in the larger periapsis precession that is seen in Fig.~\ref{spin}c.

\subsubsection{Effect of the number of whirls}

In Fig.~\ref{whirl} we show the effect of changing the number of whirls while keeping the overall orbit otherwise similar. Specifically, we make a small change in the semilatus rectum $p$, and compare three orbits with small differences in the semilatera recta
$\,\Delta(p/M)=\pm1\times 10^{-7}$. 

In all cases the rate at which the memory signal rises during the whirl part of the orbit is the same, as is seen by the indistinguishable slope in the inset in Fig.~\ref{whirl}b. Notice that also the small amplitude oscillations are nor resolvable on the scale of the figure.  Despite the indistinguishable slopes, the memory gain is substantially higher the smaller the semilatus rectum $p$. When $p/M$ is smaller the compact object executes more whirls, and although the increase in the memory effect per whirl differs by only little, the greater time spent in the whirl phase translates to a larger gain in the memory effect. 

\subsubsection{The staircase effect}

\begin{figure}[t]
\includegraphics[width=8.5cm]{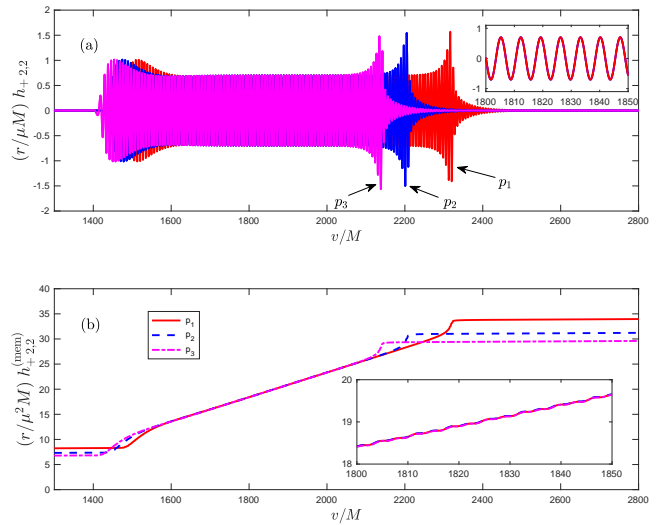}
\caption{The waveforms [upper panel (a)] and memory effect [lower panel (b)] for the $+$ polarization state of the quadrupole mode ($\ell,m=2,2$) for three zoom-whirl orbits with different semilatera recta.  We show results for $p_1/M= 2.250\,054\, 1$, $p_2/M= 2.250\,054\, 2$, and $p_3/M= 2.250\,054\, 3$ for $a/M=0.995$. Upper panel (a): The waveforms $h_{+\;2,2}$. The inset shows part of the whirl part of the waveform. On the scale of this figure, the three waveforms cannot be distinguished. Lower panel (b): The memory effect $h^{\rm{(mem)}}_{+\;2,2}$. The inset focuses on a portion of the memory signal with rapid rise. On the scale of this figure the three cases are indistinguishable.}
\label{whirl}
\end{figure}

The behavior of the memory in Figs.~\ref{spin} and \ref{whirl} is the staircase effect typical for zoom-whirl orbits: during the whirl phase of the orbital motion (and shortly afterwards) there is a sharp increase in the memory effect. When the smaller member of the system is close to apoapsis there is little contribution to the memory effect, and in the plot of the memory as a function of (retarded) time the former appears nearly flat. We can control the parameters of the staircase function: the longer the motion outside of the whirls the longer the nearly-flat portion of the steps, and the greater the number of whirls the  higher the rise in between steps. Since the memory effect is non-decreasing, during the next periapsis passage the increase pattern of the memory repeats, which gives rise to the staircase structure. 

\subsubsection{The LISA characteristic strain}

\begin{figure}[t]
\includegraphics[width=8.5cm]{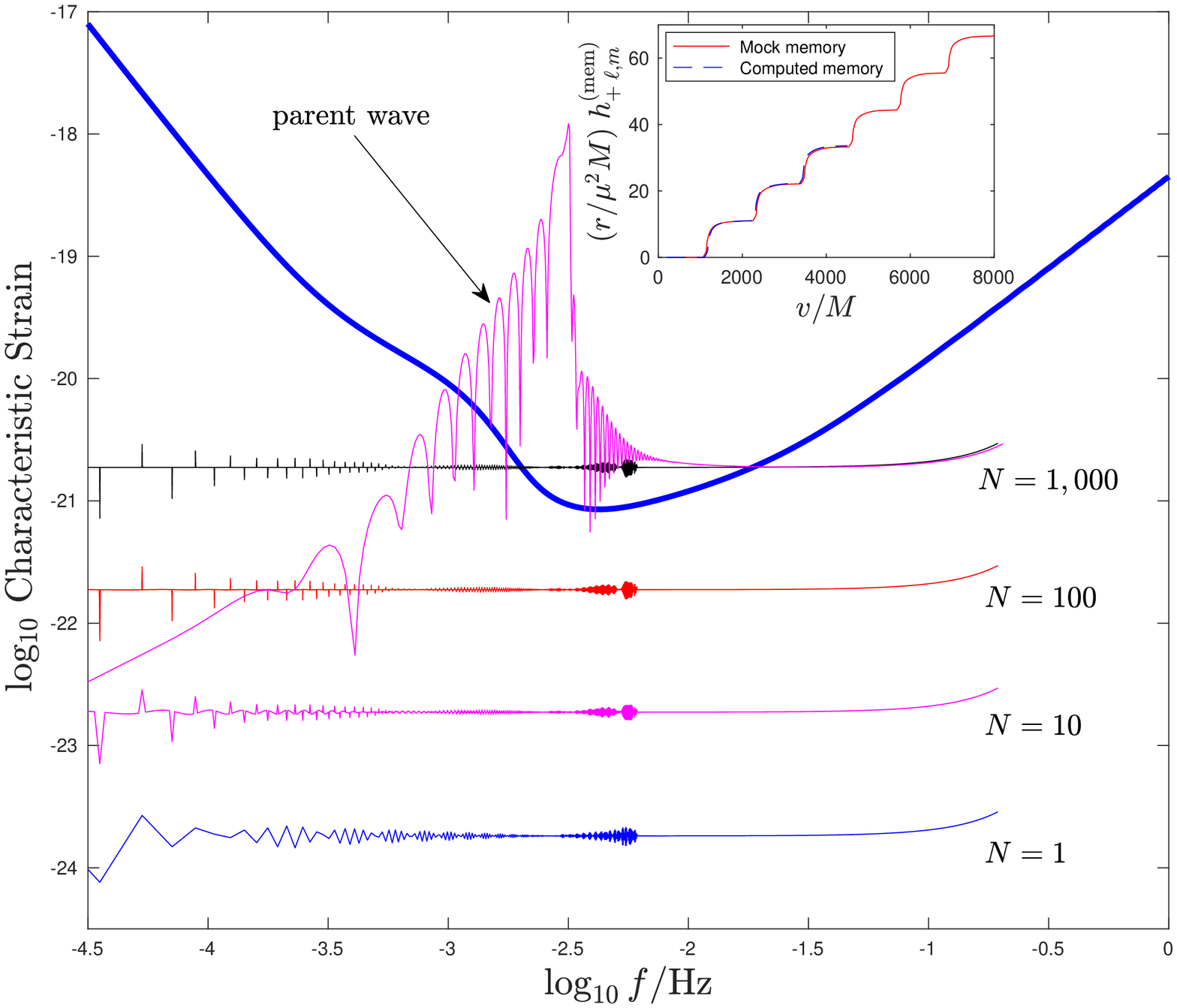}
\caption{The characteristic strain as a function of frequency for the $+$ polarized quadrupole mode ($\ell,m=2,2$)  for a zoom-whirl orbit at luminosity distance $r=100\, {\rm Mpc}$ for a central black hole of mass $M=10^7\,M_{\odot}$, spin $a/M=0.999\, 99$, and mass ratio $\mu=10^{-5}$. Shown also the characteristic strains for the memory effect with $N=1,10,100,$ and $1,000$ steps. The bold curve is the LISA sensitivity curve, obtained from \cite{robson2018,robson2018a}. The inset shows the same computed step memory as a function of (retarded) time as in Fig.~\ref{spin}, and the mock memory signal used to extrapolate the memory effect to high $N$ values. Spikes that appear in the memory strain for high $N$ values result from numerical noise associated with the construction of the mock memory.}
\label{zw-lisa}
\end{figure}

We next consider the characteristic strain of the memory effect for LISA. The short rise time implies that the memory effect could be within the frequency range for LISA. We constructed the characteristic strain using the method described in \cite{robson2018}. We show in Fig.~\ref{zw-lisa} the characteristic strain for a zoom whirl orbit shown in Fig.~\ref{spin} with central black hole of mass $M=1\times 10^7 \,M_{\odot}$ and $a/M=0.999\,99$ at luminosity distance of $100\,{\rm Mpc}$. The compact member has mass $\mathfrak{m}=1\times 10^2\,M_{\odot}$, such that the mass ratio $\mu=\mathfrak{m}/M=10^{-5}$. 

We also show in Fig.~\ref{zw-lisa} the strain for the memory effect after one, ten, a hundred, and a thousand steps of the staircase, which correspond to the respective number of periapsis passages. The evolution of very long signals is prohibitively expensive computationally. However, the memory signal as a function of (retarded) time has a very simple staircase structure. We can therefore approximate the memory of a very long signal by creating a mock memory, which is an extrapolation of the computed memory of a much shorter signal. Our mock memory retains the rise time and step duration of the computed signal, and extrapolates it to arbitrary durations. Here, we extrapolated the mock memory to $N=1,000$ periapsis passages. 
The inset in Fig.~\ref{zw-lisa} displays the computed and the mock memory signals, which on the scale of the figure cannot be distinguished. The weak dependence of the memory's characteristic strain on the frequency is a manifestation of the finite rise time of the memory signal. 

We estimate the characteristic strain  of the staircase memory signal by modeling the latter as a sum over Heaviside step functions in the short rise time model in Appendix \ref{ap}. In such a model we expect our estimate to be a good approximation in the low frequency approximation, and become inaccurate at high frequencies, because the rise time is made artificially to vanish. In the low frequency approximation we find 
the power spectral density $S_h(f)$ to be proportional to $N^2/f^2$, and the characteristic strain to be proportional to $N$. Because of the hereditary effect the characteristic strain increases linearly with the number of steps $N$. Figure \ref{zw-lisa} corroborates this expectation, and also shows weak dependence of the memory's characteristic strain on the frequency at higher frequencies, in accord with our expectation. 

However, the strain of the oscillatory parent gravitational wave is affected only little with the increase in integration time, because the time scale for changes in the parent gravitational waves, the radiation reaction time scale, is much longer than the dynamical time scale for the zoom-whirl orbits. Therefore, the memory effect strain grows in magnitude relative to the strain of the oscillatory parent signal, and with sufficiently many steps will enter the good sensitivity band of LISA. For the orbit studied in Fig.~\ref{zw-lisa}, after $N=1,000$ periapsis passages the memory's characteristic strain becomes comparable to the strain of the oscillatory parent gravitational wave.  Appropriate data analysis methods for LISA may be able to detect a strong enough memory signal from extreme mass ratio binaries. The characteristic strain of the memory effect is nearly flat with the frequency $f$, as is shown by the near constancy of the memory curves in Fig.~\ref{zw-lisa}. 

\subsubsection{Orphan Memory} 

\begin{figure}[t]
\includegraphics[width=8.5cm]{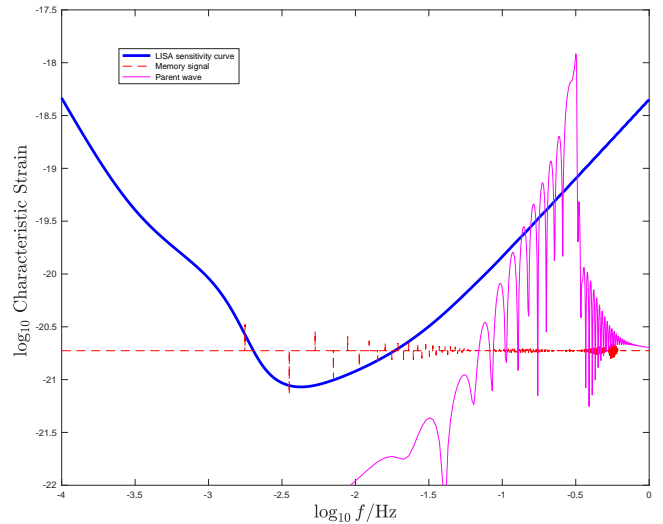}
\caption{The characteristic strain as a function of frequency for the $+$ polarized quadrupole mode ($\ell,m=2,2$)  for a zoom-whirl orbit at luminosity distance $r=100\, {\rm Mpc}$ for a central black hole of mass $M=10^5\,M_{\odot}$, spin $a/M=0.999\, 99$, and mass ratio $\mu=10^{-5}$. Shown also the characteristic strain for the memory effect with $N=1,000$ steps. The bold curve is the LISA sensitivity curve, obtained from \cite{robson2018,robson2018a}. }
\label{orph}
\end{figure}

When the oscillatory parent gravitational wave is outside the good sensitivity band for LISA, the memory signal could still be detectable \cite{mcneill_2017}. This kind of memory signal is known as orphan memory. Because of the nearly flat characteristic strain of the gravitational wave memory signal we can decrease the mass $M$ in order to shift the frequency range of the oscillatory parent signal outside of the LISA band. Specifically, in Fig.~\ref{orph} we plot the parent and memory characteristic strains as functions of the frequency for the case of a central black hole of mass $M=10^5\,M_{\odot}$ and spin $a/M=0.999\,99$, and a compact companion of mass $\mathfrak{m}=1\,M_{\odot}$ for the same orbital parameters and luminosity distance as in Fig.~\ref{zw-lisa}.

\subsection{The $\ell,m=2,0$ mode}

The $\ell,m=2,0$ mode is special, in the sense that the staircase shape has more structure. This additional structure is shown in Fig.~\ref{pluscross}a, which shows the two polarization states $h^{\rm{(mem)}}_{+,\times\;2,0}$ as functions of retarded time $v$. Each rise in a step is split into two rises and a nearly-flat section. In addition, the $\times$ polarization state is suppressed relative to the $+$ polarization state, and the memory effect is dominated by the latter. 

\begin{figure}[t]
\includegraphics[width=8.5cm]{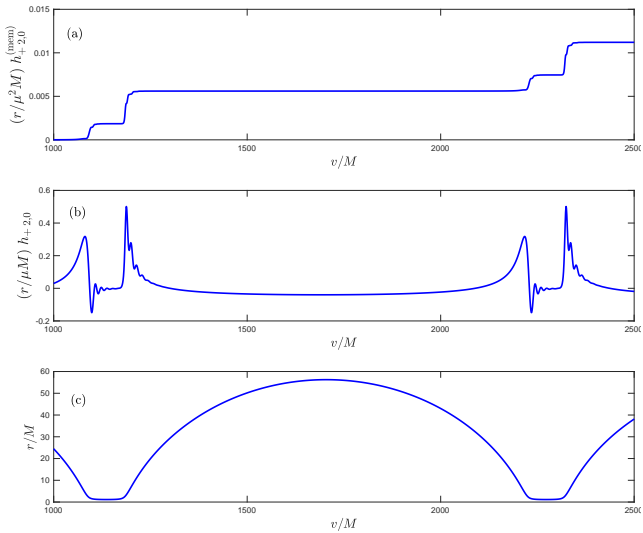}
\caption{The $\ell,m=2,0$ mode for the $+$ polarization state as a function of retarded time $v$. Panel (a): the memory effect as a function of $v$. Panel (b) The waveform. Panel (c): The zoom-whirl orbit $r$ as a function of $v$. }
\label{lm20}
\end{figure}

Figure \ref{lm20} shows the same $+$ polarization state, $h^{\rm{(mem)}}_{+\;2,0}$ as Fig.~\ref{pluscross}a together with the oscillatory parent waveform $h_{+\;2,0}$ and the radial coordinate $r$ as a function of retarded time $v$. The structure of each rise in the memory effect is associated with the whirl phase of the motion. Specifically, the $\ell,m=2,0$ does not emit oscillatory parent gravitational waves in the whirl phase of the motion, and therefore also does not contribute to the memory effect. Contributions to the memory effects are made only in the approach towards and from the whirl phase, as is shown in the inset in Fig.~\ref{pluscross}a.

\subsection{Comparing the magnitude of different modes}

\begin{figure}[t]
\includegraphics[width=8.5cm]{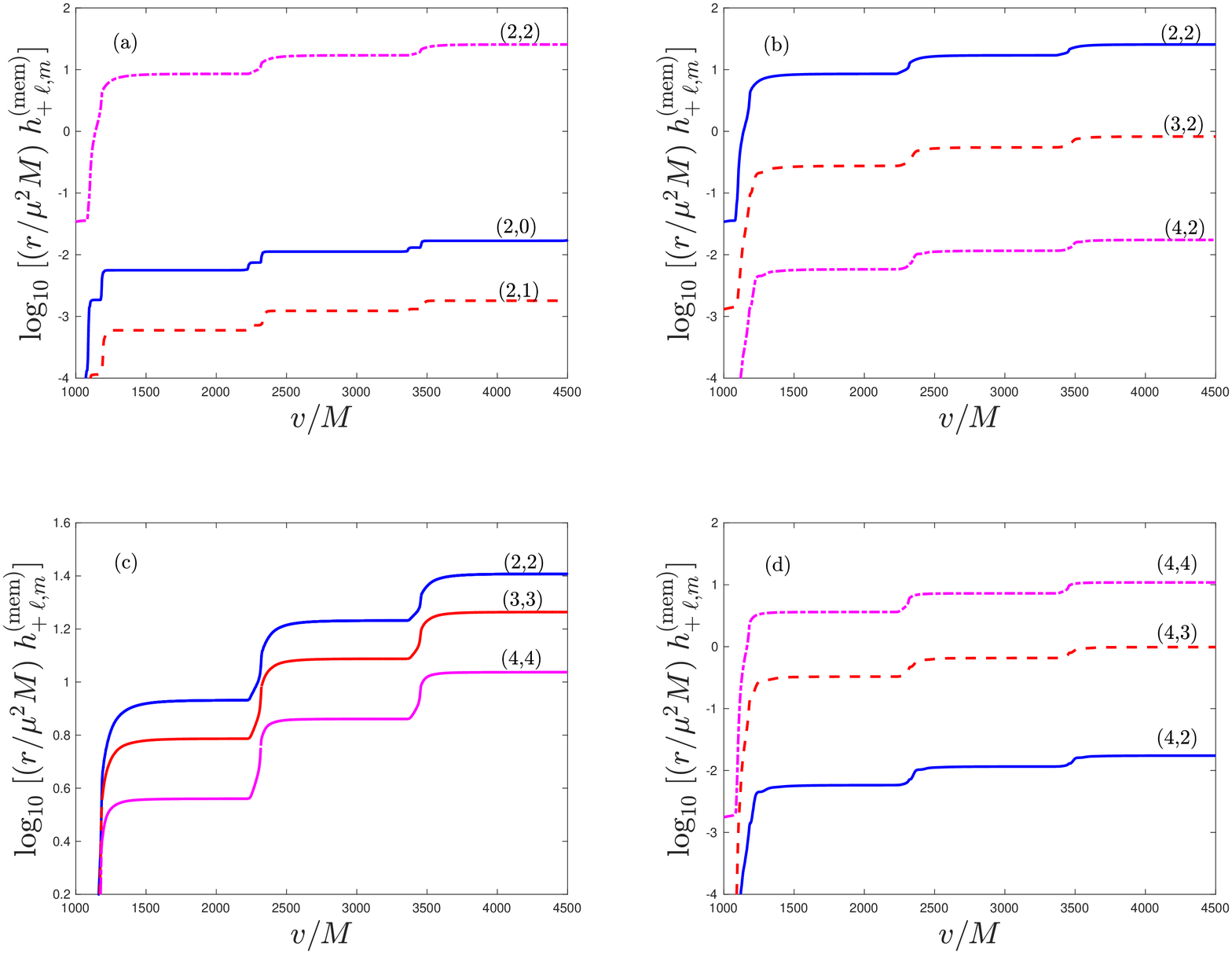}
\caption{The (logarithms of the) memory effect for the $+$ polarization state of various $\ell,m$ modes. Panel (a): The $2,m$ modes. Panel (b): The $\ell,2$ modes. Panel (c): The $m,m$ modes. Panel (d): The $4,m$ modes. We show the logarithms of the memory because of the many orders of magnitude spanned by the various modes. Here, $a/M=0.999\,9$.}
\label{lm}
\end{figure}

The relative contributions to the memory effect for various $\ell,m$ modes is shown in Fig.~\ref{lm}, which displays $h^{\rm{(mem)}}_{+\;\ell,m}$ for various groupings of modes: Fig.~\ref{lm}a shows the $h^{\rm{(mem)}}_{+\;2,m}$ modes,  Fig.~\ref{lm}b shows the $h^{\rm{(mem)}}_{+\;\ell,2}$ modes, Fig.~\ref{lm}c shows the $h^{\rm{(mem)}}_{+\;m,m}$ modes, and Fig.~\ref{lm}a shows the $h^{\rm{(mem)}}_{+\;4,m}$ modes.

The $m,m$ modes have the greatest contribution to the memory effect compared with the $\ell\ne m, m$ or the $m,n\ne m$ modes. Figure \ref{lm} suggest that in order to get a good approximation of the memory effect it may be enough to sum over the $m,m$ modes, as no other mode contributes more than $10\%$ of the contribution of the $m,m$ modes. For a more accurate calculation one would need to also include other modes to find the total effect.

\section{Extreme mass ratio inspirals}\label{sec:inspiral}

\begin{figure}[t]
\includegraphics[width=8.5cm]{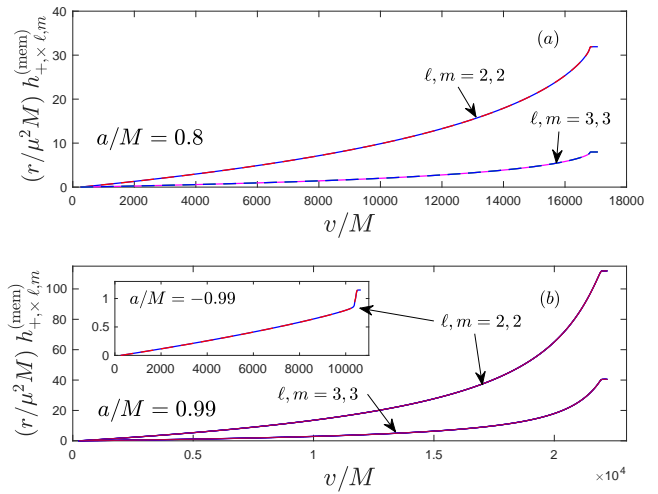}
\caption{The memory effect $h^{\rm{(mem)}}_{+,\times\;\ell,m}$ for the inspiral orbits as a function of retarded time $v$.  Upper panel (a): The case of $a/M=0.8$. Lower panel (b): The cases $a/M=0.99$ and $a/M=-0.99$ (inset). Shown are both polarization states for the $\ell,m=2,2$ and $3,3$ mode. On this scale of the figure the two polarization states cannot be resolved. The initial separations  for the three cases are as follows.  $a/M=0.99$: $r_0/M=5.025677$, $a/M=0.8$: $r_0/M=5.5198260$, and $a/M=-0.99$: $r_0/M=10.173589$, and the eccentricity $e=0$. }
\label{ins_memory}
\end{figure}

The structure of the memory signal is different for extreme mass ratio inspirals than for the zoom-whirl orbits we discussed above. We consider next equatorial inspirals for fast rotating Kerr black holes, for the prograde $a/M$ values of $0.99$ and  $0.8$, and for the retrograde value of $-0.99$. We show in Fig.~\ref{ins_memory} the memory $h^{\rm{(mem)}}_{+,\times\;\ell,m}$ for the three values of $a/M$ for the $\ell,m=2,2$ and $3,3$ modes. For these inspiral orbits we no longer see the staircase structure of the memory, but the gradual increase during the inspiral of the orbit, ending in a nearly flat maximum following merger. 

The larger the spin $a/M$ the greater the rise in $h^{\rm{(mem)}}_{+,\times\;\ell,m}$. This effect can be explained by the longer evolution and larger number of cycles executed because the innermost stable circular orbit is closer to the black hole's event horizon the larger $a/M$  \cite{bardeen72}.  In the retrograde case the innermost stable circular orbit is farther out, which explains the smaller number of cycles and therefore the smaller accumulation of the memory effect.  

\begin{figure}[t]
\includegraphics[width=8.5cm]{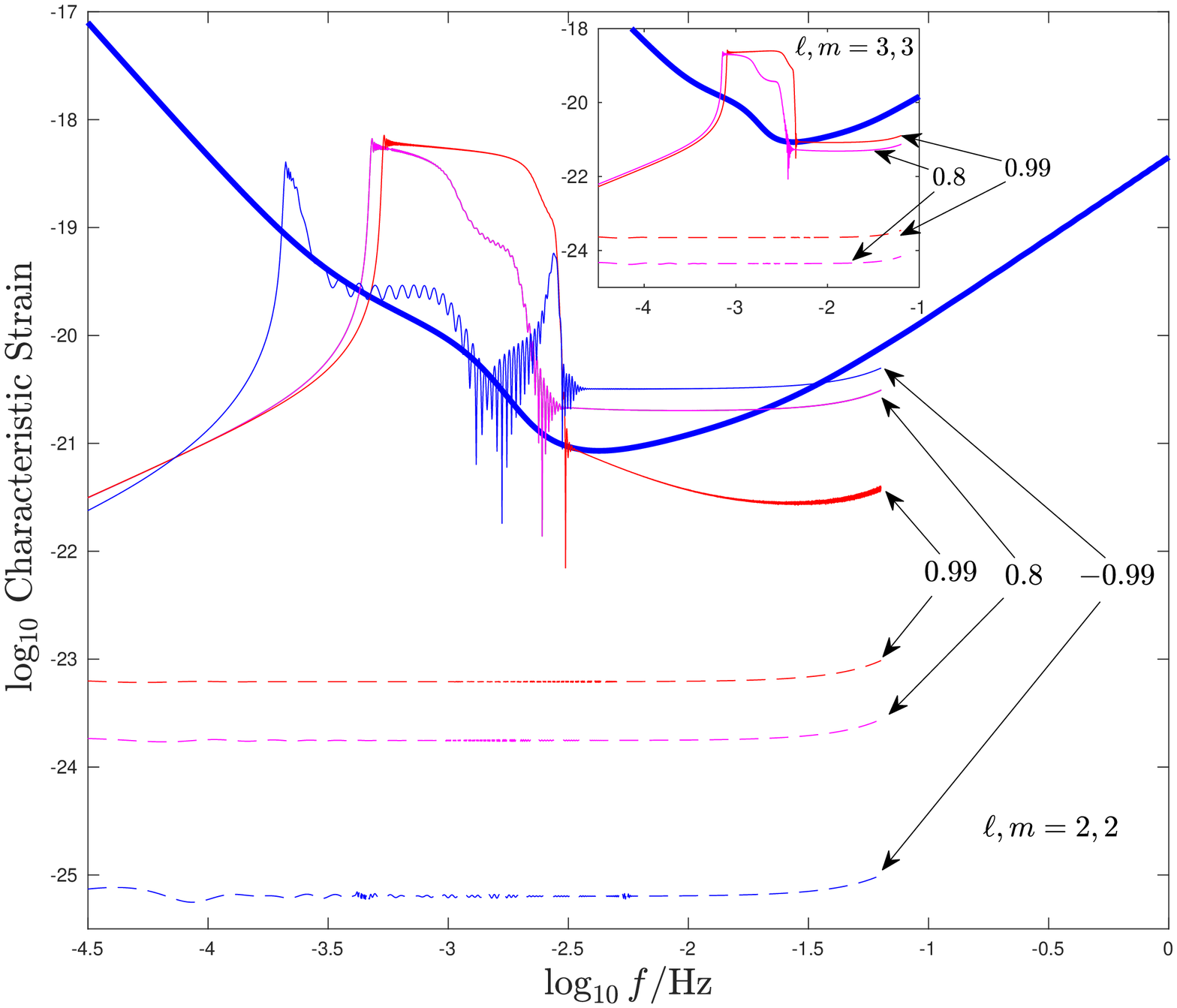}
\caption{The characteristic strain (solid curves) as a function of frequency for the $+$ polarized quadrupole mode ($\ell,m=2,2$)  and for the $\ell,m=3,3$ mode (inset) for inspiral orbits at luminosity distance $r=100\, {\rm Mpc}$ for a central black hole of mass $M=10^7\,M_{\odot}$ and mass ratio $\mu=10^{-5}$. The spin values are $a/M=0.99$, $0.8$, and $-0.99$.  
Shown also are the corresponding characteristic strains for the memory effect (dashed curves). The bold curve is the LISA sensitivity curve, obtained from \cite{robson2018,robson2018a}. }
\label{ins_sensitivity}
\end{figure}

Despite the different memory signals, the characteristic strains behave similarly in the two cases, and is only very weakly dependent on the frequency $f$ for both the $\ell,m=2,2$ and $3,3$ modes, as is shown in Fig.~\ref{ins_sensitivity}. Unlike the zoom whirl case, for inspirals the memory signal is below the LISA sensitivity curve, and consequently the chances of detection are strongly restricted. Figure \ref{ins_sensitivity} suggests that it could be easier to detect the memory effect from extreme mass ratio inspirals for higher prograde spin systems than from lower prograde or retrograde spin systems.

\section{Concluding remarks}\label{sec:conc}

The memory effect from zoom-whirl orbits of extreme mass ratio systems is intriguing from the observational point of view because it increases linearly in magnitude with the periapsis passages, whereas the oscillatory parent wave does not. Therefore, its chances of detection increase with observation time, but appropriate data analysis methods would need to be developed. That is not the case for inspiral memory, where we no longer have the advantage of the staircase structure of the memory signal. We find that with sufficiently many periapsis passages the memory signal may enter the good sensitivity band of LISA. 

Our numerical experiments show sensitivity of the memory effect to the systems orbital parameters. Accurate measurement of the memory effect could therefore potentially be used for precision parameter estimation.  This dependence of the memory effect on the parameters of the system is limited to the magnitude of the effect, but not to its accumulation rate. The latter depends on the whirl parameters. 

In Fig.~\ref{zw-lisa}  we used favorable parameters: The sources were taken at a relatively short distance (100 Mpc), the spin $a/M$ was taken to be large, the mass of the central black hole was taken to be high ($10^7\,M_{\odot}$), and the mass ratio was taken to be small ($\mu=10^{-5}$) so that the memory effect would appear significant with just a thousand periapsis passages. For many systems the parameters would not be as favorable. However, when more periapsis passages are considered, the memory effect could still be observable with appropriate search and data analysis techniques. 

We present results for equatorial orbits that are aligned such the memory effect is maximized ($\iota=\pi/2$). When generic, non-equatorial orbits are considered, the inclination factor $\Phi(\iota)$ would suppress the memory effect accordingly during parts of the orbit, and would create -- in conjunction with the more intricate parent waveform -- a more detailed memory curve than the simple staircase structure we show here. 

In all the cases we considered we found that the characteristic strain for the memory signal is only weakly dependent on the frequency.  This property of the gravitational wave memory effect may have observational consequences. It is conceivable that with a sufficiently large number of periapsis passages the low frequency part of the memory signal could be detected by a pulsar timing array, and ever that the memory of a LISA signal would be detectable by a pulsar timing array. It is likewise conceivable that the memory of a wave detected by a terrestrial detector could be detectable by LISA. 

We presented separately the results for each $\ell,m$ harmonic mode. To obtain the total memory effect one should sum up over all such modes. 

We used a simple kludge approximation to calculate the memory effect, which is based on the approach of \cite{johnson19}. More accurate approximation methods, such as those discussed in Section \ref{sec:model} and specifically the method of \cite{talbot2018}, could estimate the actual memory effect more accurately, but we expect that the major conclusions here would not be affected by such a more robust approach. 

The memory effect increases with the spin of the central black hole, which would make their memory effect more detectable. Therefore, if existing, nearly extreme black holes could have an additional observational signature \cite{wiggles,burko-khanna-16,transient}, that would be biased in search results.

\section*{Acknowledgements}  
The authors are thankful to Nur Rifat for generating the LISA sensitivity curve, and to Neil Cornish, Tousif Islam, Aaron Johnson, and Dan Kennefick for discussions. Many of the computations were performed on the MIT/IBM Satori GPU supercomputer supported by the Massachusetts Green High Performance Computing Center (MGHPCC). G.K.~acknowledges research support from NSF Grants No. PHY-1701284 and No. DMS-1912716 and 
Office of Naval Research/Defense University Research Instrumentation Program
(ONR/DURIP) Grant No. N00014181255.

\appendix
\section{Estimating the characteristic strain for a staircase signal}\label{ap}
We estimate the expected amplitude spectral density in the short rise time limit. Specifically, we model the memory signal to be of the general form 
\begin{equation}
h^{\rm{(mem)}}\approx \sum_{k=0}^{N-1}\,\delta h_k\;\Theta(v-kT)\, ,
\end{equation}
where $\Theta(v-v_k)=1$ for $v>v_k$ and $0$ otherwise is the Heaviside step function, and $\,\delta h_k$ is the rise of each step. Here, $T$ is the duration of the flat part of the step. Since the orbital evolution is on a much longer time scale than the dynamical time scale for each step, we take $\,\delta h_k \approx \,\delta h$ to be independent of $k$, such that $h^{\rm{(mem)}}\approx \,\delta h\;\sum_{k=0}^{N-1}H(v-kT)$. 

Because of the linearity of the Fourier transform, ${\cal F}_v[\sum_{k=1}^{N-1}\Theta(v-kT)](f)=\sum_{k=1}^{N-1}{\cal F}_v[\Theta(v-kT)](f)$. Recall that ${\cal F}_v[\Theta(v-kT)](f)=\frac{e^{-kfT}}{2\pi i f}$ (for $f\ne 0$). 
Therefore, the amplitude spectral density 
\begin{eqnarray}
A(f)&=&{\cal F}_v\left[h^{\rm{(mem)}}(v)\right](f)\approx \frac{\,\delta h}{2\pi i f}\,\sum_{k=0}^{N-1} e^{-ikfT}\nonumber \\
&=& \frac{\,\delta h}{2\pi i f}\,\frac{1-e^{-iNfT}}{1-e^{-ifT}}\, .
\end{eqnarray}

The power spectral density is 
\begin{equation}
S_h(f)=\frac{\left|A(f)\right|^2}{2\mathcal{T}}=\frac{(\,\delta h)^2}{8\pi^2f^2{\mathcal T}}\,\frac{\,\sin^2\left(\frac{NfT}{2}\right)}{\,\sin^2\left(\frac{fT}{2}\right)}\, ,
\end{equation}
where ${\mathcal T}$ is the observation time. We point out in passing the similarity between our calculation and the Fraunhofer diffraction of a diffraction grating. 
In the low frequency approximation, $f\ll 2/T$, we find
\begin{equation}
S_h(f)\sim\frac{N^2\,(\delta h)^2}{8\pi^2f^2{\mathcal T}}\, .
\end{equation}

The characteristic strain is then 
\begin{eqnarray}
h_{\rm eff}(f)&=&\frac{4\sqrt{2}}{\sqrt{5}}f\,{\mathcal T}^{1/2}\,S_h^{1/2}(f)\nonumber \\
&\sim& \frac{2\,N}{\sqrt{5}\pi}\,(\,\delta h)\, .
\end{eqnarray}

To estimate the approximations we made, we take $M=10^7\,M_{\odot}$, $\mu=10^{-5}$, $a/M=0.999\,99$, at luminosity distance $r=100\,{\rm Mpc}$. We read off $\,\delta h\sim 10\,\mu^2M/r$ from Fig.~\ref{spin} for the zoom-whirl orbit therein. For just one step, $N=1$, we find
$\left.h_{\rm eff}(f)\right|_{N=1}\approx 4\times 10^{-24}$, while for a thousand steps, $N=1,000$ we find $\left.h_{\rm eff}(f)\right|_{N=1,000}\approx 4\times 10^{-21}$, close to the results shown in Fig.~\ref{spin}.

\end{document}